\documentclass[a4paper]{article}

\usepackage{INTERSPEECH2022}
\usepackage{amsmath,graphicx}
\usepackage{multirow}
\usepackage{multicol}
\usepackage{makecell}
\usepackage{color,xcolor}
\usepackage{caption}
\usepackage{subcaption}
\title{DEFORMER: Coupling Deformed Localized Patterns with \\Global Context for Robust End-to-end Speech Recognition}
\name{Jiamin Xie$^1$, John H.L. Hansen$^1$}
\address{
  $^1$ Center for Robust Speech Systems (CRSS), University of Texas at Dallas, TX, 75080}
\email{\{Jiamin.Xie, John.Hansen\}@utdallas.edu}

\begin{document}
\maketitle
\begin{abstract}
Convolutional neural networks (CNN) have improved speech recognition performance greatly by exploiting localized time-frequency patterns. But these patterns are assumed to appear in symmetric and rigid kernels by the conventional CNN operation. It motivates the question: What about asymmetric kernels? In this study, we illustrate adaptive views can discover local features which couple better with attention than fixed views of the input. We replace depthwise CNNs in the Conformer architecture with a deformable counterpart, dubbed this ``Deformer". By analyzing our best-performing model, we visualize both local receptive fields and global attention maps learned by the Deformer and show increased feature associations on the utterance level. The statistical analysis of learned kernel offsets provides an insight into the change of information in features with the network depth. Finally, replacing only half of the layers in the encoder, the Deformer improves +5.6\% relative WER without a LM and +6.4\% relative WER with a LM over the Conformer baseline on the WSJ \textit{eval92} set.
% Finally, we test the generalization of the proposed architecture on the SWB spontaneous telephony speech corpus.
%  The generalization performance of the model on the Hub5'00 benchmark provides a ~\% WER using a LM.
% Through studies on the WSJ corpus, we compare the Deformer to the Conformer. Both localized and global patterns discovered .
\end{abstract}

\section{Introduction}
Convolution neural networks (CNN) are widely used to process signals generated in various domains, including images \cite{krizhevsky2012imagenet, chollet2017xception,he2016deep}, languages \cite{gehring2016convolutional,conneau-etal-2017-deep,jiang2020convbert}, and sounds \cite{zhang2015robust,yang2017midinet, kothapally2022skipconvgan}. The basic operation of the CNN computes a weighted sum of the input within a kernel. The kernel shape is defined by the kernel size and dilation parameter, which controls the gap between consecutive positions inside a kernel. For example, a grid-like kernel has dilation of 1.  The kernel sweeps the input signal yielding output at each location sequentially. Different from the convolution used in signal processing \cite{dimitriadis2005robust,long1996wavelet}, the CNN uses learnable filters and does not flip the input. Modern methods stem from the basic CNN operation, including dynamic convolution \cite{chen2020dynamic}, separable convolution \cite{hu2018squeeze}, differential strides \cite{riad2022learning}, etc. The kernel shape, however, is commonly kept symmetric and rigid.
% mehrotra1992gabor,zetzsche1989invariant

In speech recognition, CNNs are applied directly or modified for both acoustic modeling in hybrid systems and end-to-end (E2E) systems. In \cite{von2019multi, han2021multistream}, multiple streams of CNNs with varying dilation rates and kernel sizes were combined to extract acoustic features at different input resolutions. For E2E systems, the VGG-like convolutional network is one popular choice used to subsample and encode speech \cite{zhang2017very,lu2020exploring}. In \cite{li2019multi}, a multi-encoder structure takes further advantage of architectural differences in addition to multi-resolution. Recent advancements are inspired by works in the vision community. QuartzNet \cite{kriman2020quartznet} reduces model parameters using separable convolution \cite{hu2018squeeze}. ContextNet \cite{han2020contextnet} utilizes squeeze-and-excitation \cite{hu2018squeeze} to channel-wise modulate the output by pooling a global context from the full utterance.
The Conformer architecture \cite{gulati2020conformer} further integrates both convolution and attention \cite{vaswani2017attention} to couple both local and global relationships, which is found to be useful in many speech related tasks \cite{chen2021continuous,ma2021end}.
%,narayanan2021cross
\begin{figure}[h!]
  \centering
  \includegraphics[width=0.9\linewidth]{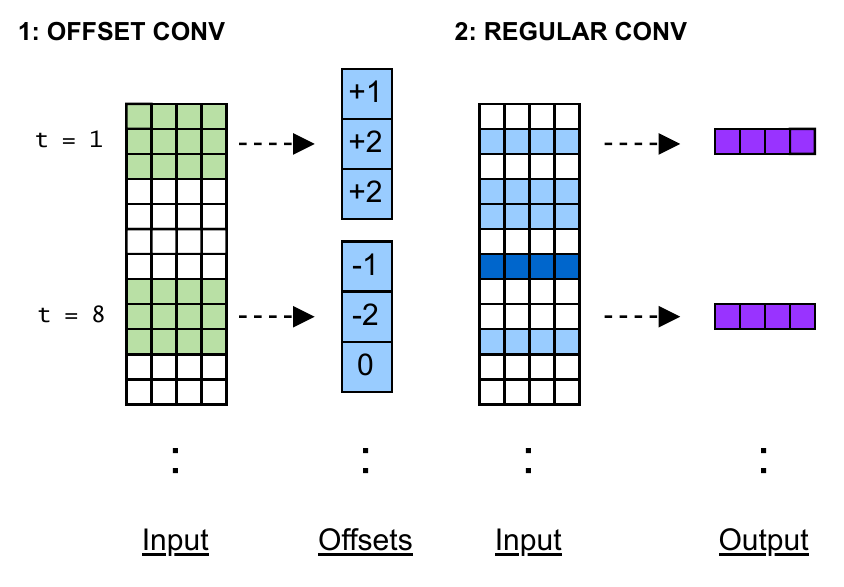}
  \caption{Deformable convolution steps, shown with kernel size=3, stride=7, dilation=1. Note at t=8, the kernel positions become overlapped after the offset operation.}
  \label{dcnn_illustration}
\end{figure}

The wide use of CNNs in speech recognition systems suggests the importance of localized patterns. One speech recording can capture both task-relevant and task-irrelevant information (e.g. speech and speaker vs. noise and channel). But all regions are scanned and processed \textit{equally} when a convolution kernel slides across the input audio. This weakens a model’s capacity to capture localized patterns because the kernel weights have to adjust and de-emphasize irrelevant contents. Thus, if a model can constantly adapt to the input by predicting task-relevant regions, the convolution can learn better weights to strengthen the information in a local context. This fundamental problem is considered using the deformable CNN \cite{dai2017deformable}, originally proposed to operate on 2-D input for image segmentation tasks. In \cite{an2021deformable}, the 1-D version is first used for speech recognition. The idea was to scan the input and predict offsets that deform the kernels in a subsequent convolution. As illustrated in Fig. \ref{dcnn_illustration}, one deformable convolution consists of an offset CNN and an output CNN. Similar to CNNs, the deformable CNN employs parameters such as kernel size, stride, and dilation. But unlike conventional CNNs, the hyperparameters only specify the input sequence locations of the offset CNN. The input sequence locations of the output CNN are deformed by the predicted offsets. Therefore, a deformable CNN can help build localized patterns by focusing on the most relevant information.
\begin{figure*}[h!]
  \centering
  \includegraphics[width=0.9\textwidth]{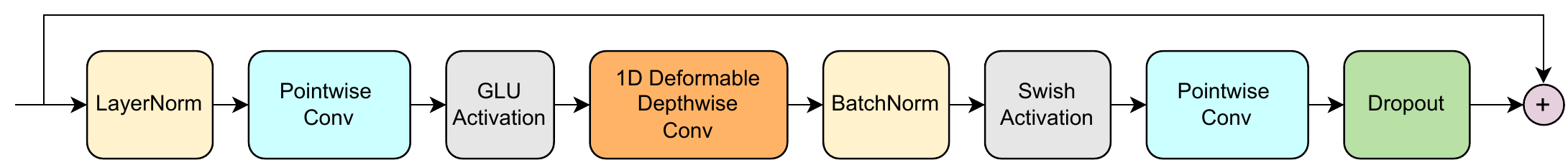}
  \caption{Deformable convolution module: following the GLU-activated pointwise convolution units, the 1-D depthwise deformable convolution is implemented to replace the regular 1-D depthwise convolution.}
  \label{fig:deformer_arch}
  \vspace{-1em}
\end{figure*}

In this study, we present a novel encoder model called Deformer to learn deformed localized patterns. We introduce the deformable variant of the depthwise CNN that the encoder uses. Lastly, we provide an in-depth analysis of the patterns and the deformation learned. The paper is organized as follows. Sec. 2 discusses related work. Sec. 3 outlines the Deformer encoder design. Sec. 4 presents the analysis of offsets and attention patterns as well as visualizations. Sec. 5 lists the experimental setup and illustrates the results and our findings. Sec. 6 concludes this paper and provides an outlook for future work. We thank our collaborators and acknowledge them in Sec. 7.

\section{Related Work}
The applications of the deformable CNN focused on areas such as video processing \cite{xu2019learning} and speaker recognition on the spectrograms \cite{zhang2020speaker}. The 1-D deformable CNN was first proposed for action segmentation in \cite{lei2018temporal}. It was only recently that An et al. applied the deformable CNN to speech recognition \cite{an2021deformable}. In detail, they modified a 7-layer standard TDNN model by making the last several layers deformable. Compared with the base TDNN, the introduction of deformed kernels has shown superior performance on both time-shifted and regular speech. Using neural architecture search, the deformable TDNN achieved a remarkable 10\% relative WER improvement on the \textit{eval92} set of WSJ over the base TDNN. Their work designed a system based on the CTC-CRF-based approach \cite{xiang2019crf} to end-to-end (E2E) ASR. Our study here differs by focusing on an encoder design for LAS-based \cite{chan2015listen} E2E ASR systems, which works with a deep architecture and attention mechanisms.

% There are works of using 2D deformable CNN on spectrogram for speaker recognition \cite{zhang2020speaker} and video processing \cite{gao2022multiscale}. \JM{TODO: fill in more in this paragraph}
% Therefore, we aim to study the effectiveness of asymmetric kernels by applying deformable CNN to convolution models for speech recognition. We further study how well local patterns learned by deformable CNN may change the learning of attention maps (i.e. global patterns). \JM{TODO: fill in more in this paragraph}

\section{Deformer Encoder}
 Our proposed Deformer encoder is similar to the Conformer encoder, which has the SpecAug \cite{park2019specaugment}, convolution subsampling layers, and stacked attention-convolution blocks. The Deformer uses a deformable convolution module, as depicted in Fig. \ref{fig:deformer_arch}. 
% \subsection{Deformable Subsampling Layers}
% The deformable subsampling layers are a natural modification using 2D deformable CNNs. 
% \subsection{Deformable Convolution Module}
\subsection{1-D Deformable Convolution}
% maybe change to associate with figure 1
The \textit{1-D} deformable convolution \cite{an2021deformable} is built on regular convolutions and contains three steps, 1) offset convolution, 2) linear interpolation, and 3) output convolution. The first step uses convolution to predict input position offsets $\Delta p \in {\rm I\! R^{T\times K}}$, where $T$ is the input length and $K$ is the kernel size. The prediction uses the initial input positions $p_{0} \in {\rm I\! R^{T\times K}}$, which are determined by the kernel size, stride, and dilation parameters. These are positions shown green in Fig \ref{dcnn_illustration}. Given an input sequence $X$, the first step computes the offset sequence $\Delta p$ by, 
\begin{equation}
    \Delta p = Conv1D_{offset}(X,\ p_{0}).
\end{equation}
Since the offset values $\Delta p$ are fractional, the second step will produce a linear interpolation of the input to get corresponding input values at $\Delta p$. This procedure is computed by,  
\begin{equation}\label{p_hat}
    p' = p_{0}+\Delta p,
\end{equation}
\begin{multline}
    X(p') = X(\lfloor p' \rfloor) * (\lfloor p' \rfloor - p' + 1 ) \\ 
    + X(\lfloor p' \rfloor + 1) * (p' - \lfloor p' \rfloor),
\end{multline}
where $\lfloor.\rfloor$ is the floor operation and $p'$ is the position sequence after deformation, as shown blue in Fig \ref{dcnn_illustration}. Lastly, the output convolution is applied to the input at deformed locations. For an output sequence $Y$, this is computed by,
\begin{equation}
    Y = Conv1D_{output}(X,\ p'). \\
    \label{eq4}
\end{equation}
Note when $\Delta p$ is set to zero, we revert back to the regular \textit{1-D} convolution that uses the same kernel size, stride, and dilation parameters as which produced the positions $p_{0}$.

\subsection{1-D Deformable Depthwise Convolution}
Consider 2-D input $X \in {\rm I\! R^{T\times F}}$ with time and frequency information, the depthwise operation slices the input along the $F$ dimension into distinct groups. For a 1-D depthwise convolution, each input slice $X_{g} \in {\rm I\! R^{T\times F/g}}$ is then convolved with a shared weight matrix to produce an output slice $Y_{g} \in {\rm I\! R^{T\times N/g}}$, where $N$ is the output dimension size and $g$ is the number of groups. The idea is that the spatial ($T$) and channel ($F$) information of the input can be separated \cite{chollet2017xception}.
Since the \textit{1-D} deformable convolution deforms the input spatial locations, we make the \textit{1-D} depthwise convolution in the Conformer deformable. This means making the $Conv1D_{output}$ in Eq. \ref{eq4} a depthwise operation. It is also intuitive to make the $Conv1D_{offset}$ depthwise by setting the \textit{deformable groups}, which removes channel dependencies across groups. However, our empirical result suggests using the same group of offsets for all channels outperforms different groups, as further explained in section 6.2.

\section{Pattern Analysis}
We conduct a statistical analysis of offset values learned in each deformable layer to interpret deformation. We visualize both localized and global patterns obtained. The spread of attention distribution is evaluated using metrics from \cite{yang2020understanding}.
\subsection{Setup}
The best-performing model on the combined WSJ development and evaluation set is used for analysis. Note we have only used development set in parameter selection. The model contains a 12-layer Deformer encoder and a 6-layer Transformer decoder, predicting the probability of letters. The encoder consists of a mix of deformable and non-deformable CNN layers. Table \ref{config} lists the detailed configurations of our encoder network. 
\begin{table}[h!]
\centering
\resizebox{\columnwidth}{!}{
    \begin{tabular}{c c c}
    \toprule
    \textbf{Configurations} & \textbf{Deformable Layers} & \textbf{Non-deformable Layers} \\
    \midrule
    Layer Index & \{1, 6, 7, 10, 11\} & \{0, 2-5, 8, 9\}\\
    Layer Dimensions & 256 & 256 \\
    Attention Heads & 4 & 4 \\
    Kernel Size & 15 & 15 \\
    Dilation & 1 & 1 \\
    Stride & 1 & 1 \\
    Convolution Groups & 256 & 256 \\
    Deformable Groups & 1 & 1 \\
    \bottomrule
    \end{tabular}
}
\caption{Deformer Encoder Configuration}
\label{config}
\vspace{-2.2em}
\end{table}

\begin{figure*}[h!]
    \centering
    \begin{subfigure}[b]{0.24\textwidth}
        \centering
        \includegraphics[width=\textwidth]{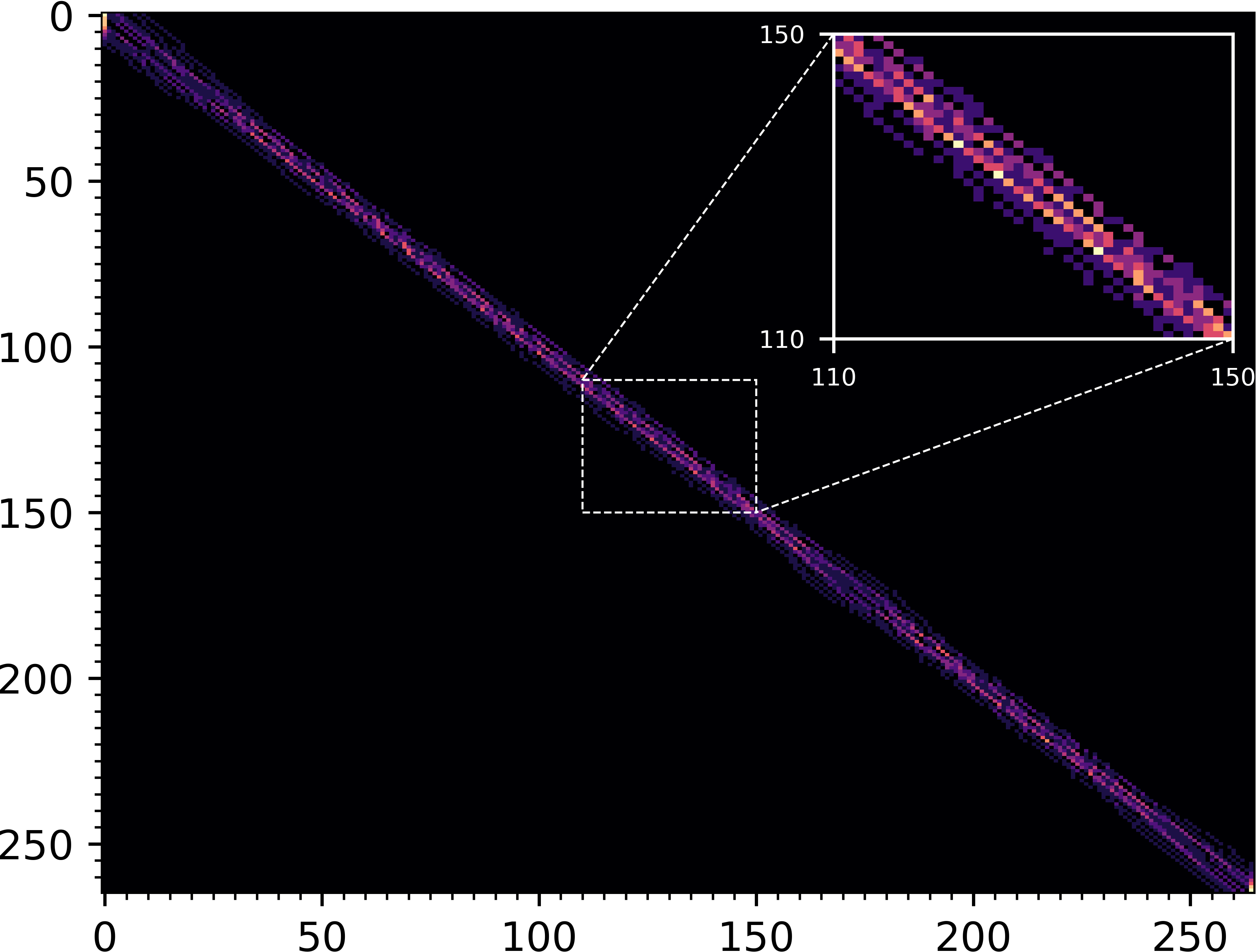}
        \caption{Deformer Layer 1}
        \label{fig:five over x}
    \end{subfigure}
    \begin{subfigure}[b]{0.24\textwidth}
        \centering
        \includegraphics[width=\textwidth]{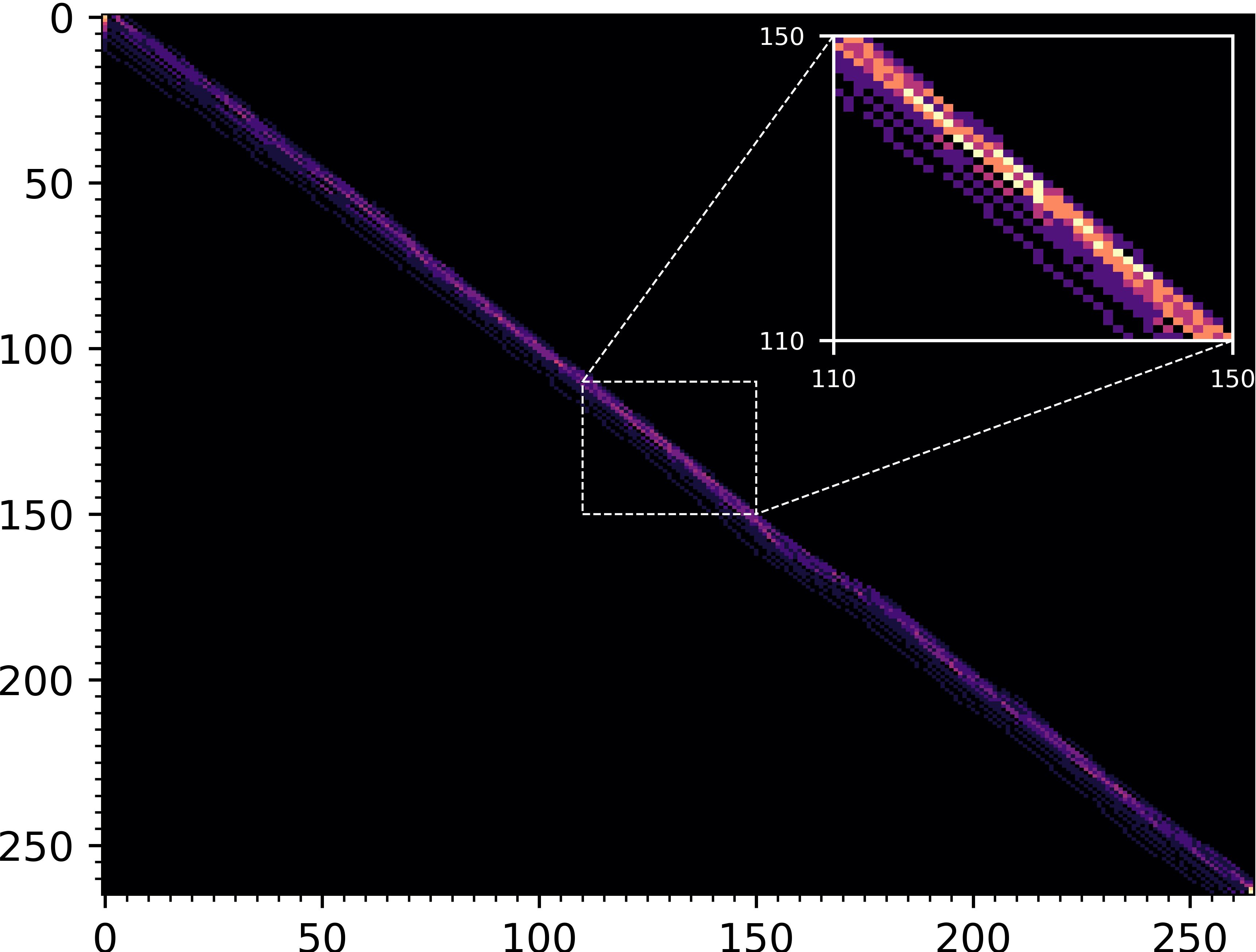}
        \caption{Deformer Layer 6}
        \label{fig:three sin x}
    \end{subfigure}
    \begin{subfigure}[b]{0.24\textwidth}
        \centering
        \includegraphics[width=\textwidth]{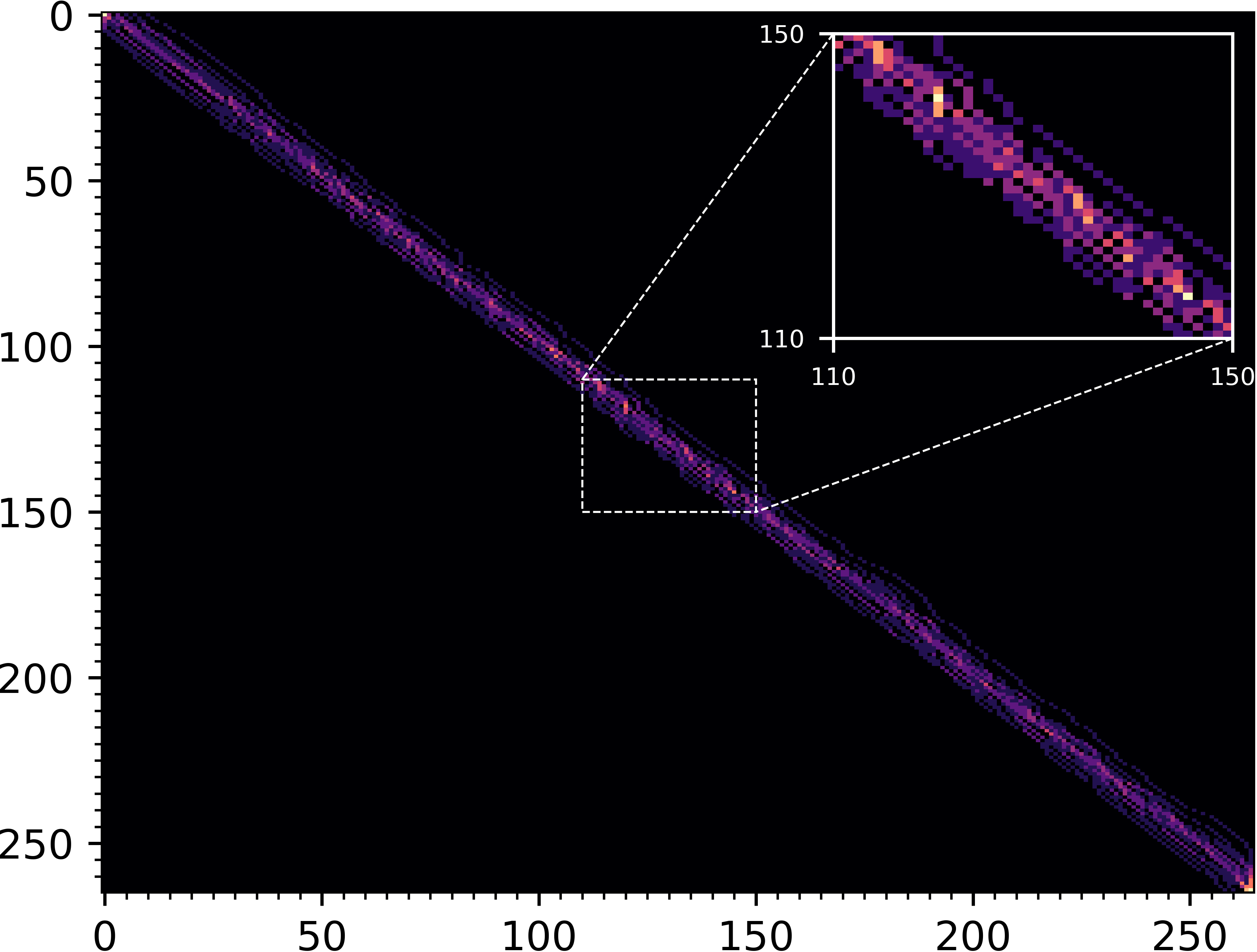}
        \caption{Deformer Layer 7}
        \label{fig:five over x}
    \end{subfigure}
         \begin{subfigure}[b]{0.24\textwidth}
        \centering
        \includegraphics[width=\textwidth]{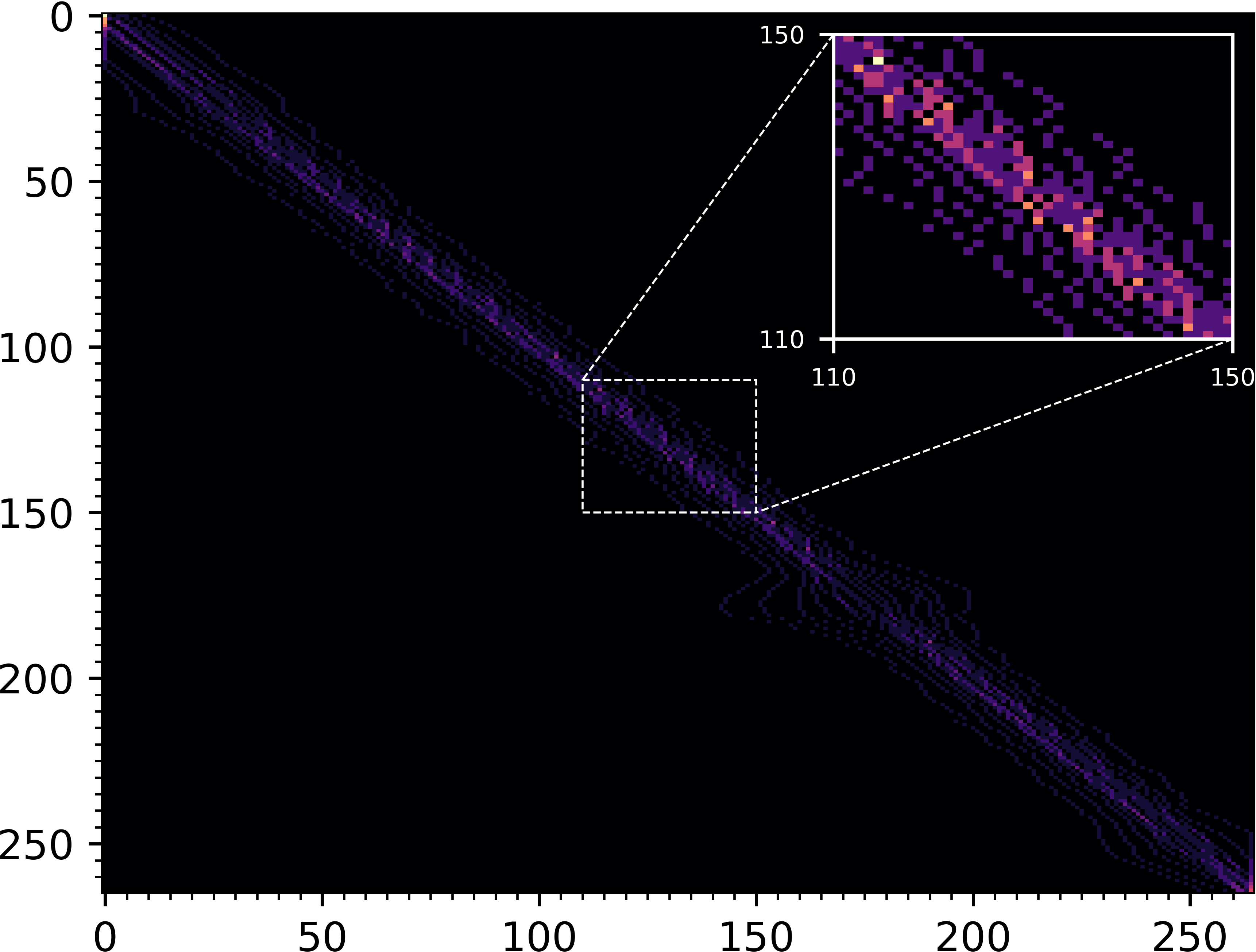}
        \caption{Deformer Layer 11}
        \label{fig:1a}
    \end{subfigure}
        \caption{Unrolled convolution locations with deformed kernels on decoding the 4k0c030i utterance. A section is zoomed in and displays a kernel pattern on the upper right corner. The color intensity reflects the number of overlaps at each location (higher the brighter). The horizontal axis shows the input positions and the vertical axis shows the output positions.}
        \vspace{-1em}
        \label{local_patterns}
\end{figure*}
\subsection{Localized Patterns and Offset Statistics}
\vspace{-0.1em}
Fig. \ref{local_patterns} unrolls the deformable convolution and visualizes the learned receptive fields of kernels after the offset. From the utterance perspective, the network learns to place kernels along the diagonal, which affirms a monotonic alignment is optimal for speech recognition. In each row, the receptive field of a kernel does not show a large deviation from the local context even though we do not limit the magnitude for offsets (except to the boundaries of the utterance). Hence, localized patterns are beneficial overall despite a few exceptions in deep layers (e.g. positions from (165,165) to (185,185) in Fig. 5d). It is worth noting that the addition of offset can make several positions \textit{overlap}. By observing the overlaps, we conclude a concentration pattern adopted by the network. The center of a kernel has a high focus on the local information, which shows typically bright-colored. The sides of a kernel scatter to distant information that is only supportive, which appear light-colored. Thus, we think the deformable CNN constructs meaningful features by shifting to locations where it helps enhance a locally present knowledge.
\vspace{-0.5em}
% the observation of regions with high color intensity reveals a concentration-like pattern inside a deformed kernel.
\begin{figure}[h!]
  \centering
    \includegraphics[scale=0.4]{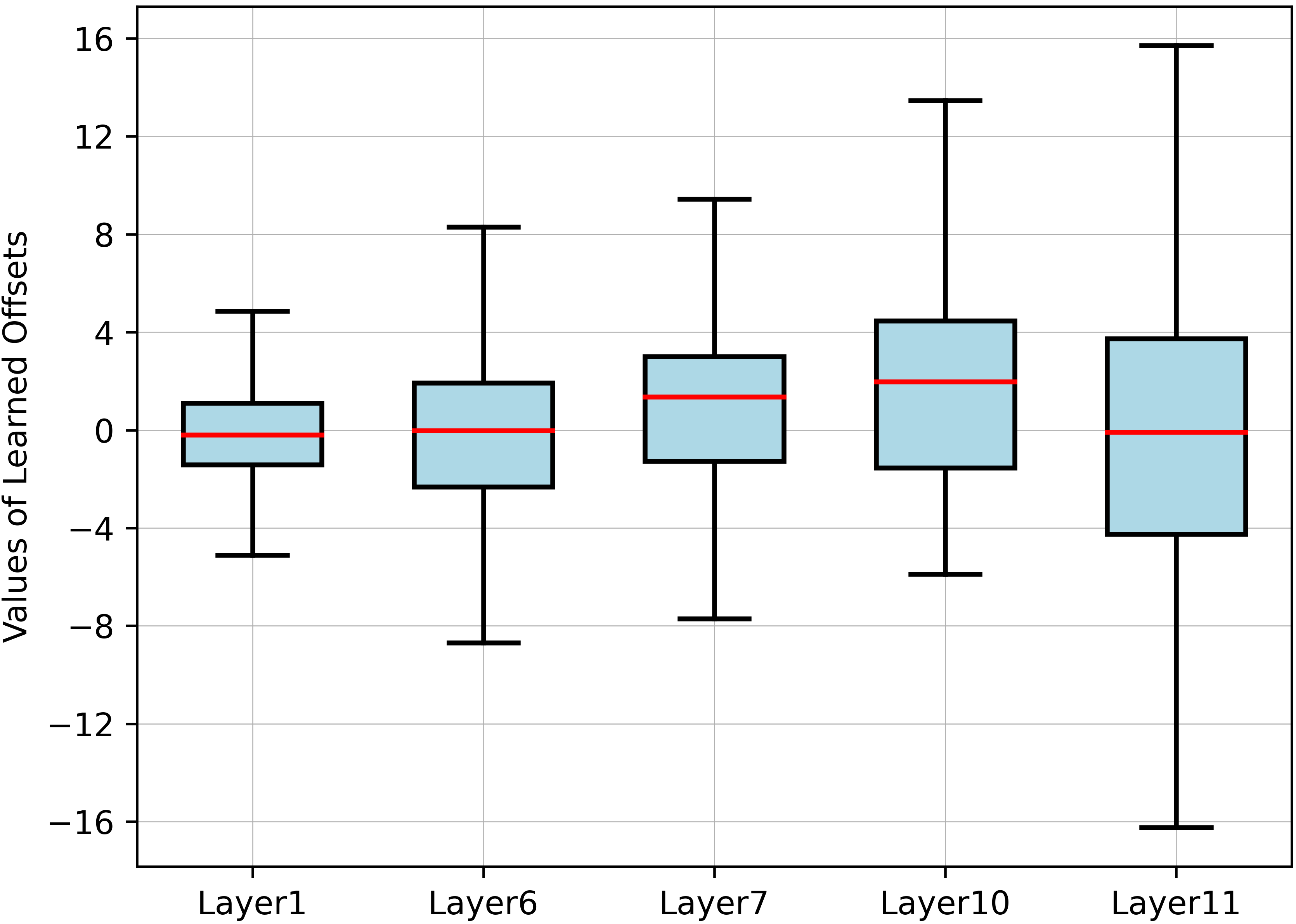}
  \caption{Boxplots of offset distribution in each deformable layer on the combined dev93 and eval92 set. The box boundaries are the first (Q1) and the third quartiles (Q3). The red line is the median. The whiskers denotes values within a 1.5 times interquartile range below Q1 and above Q3.}
  \label{boxplot}
\end{figure}
\vspace{-2em}
%TODO: fix Encoder to Layer ..
%TODO: Details in caption  .. and zoom
%TODO: Use same utterance, and mention above across utterance distribution
%TODO: Talk more on discovered patterns
%TODO: Find better color ..

Since offset values change with different input, Fig. \ref{boxplot} displays the distributions of offsets collected on the corpus level. We first observe a larger receptive field obtained in deeper layers, shown by an increasing spread of the distribution. Next, we confirm the concentration pattern persists across different utterances (as previously discussed on a single utterance). Each set of Q1 and Q3 values in Layers 1, 6, and 11 is nearly symmetric around the zero medians. This explains the deformation around the kernel center because the focus (overlap) requires shifting nearby positions either to its past or future. The distributions also present a long tail. This explains the deformation of the kernel sides, where supporting information is afar and scattered. 
% In summary, it suggests that similar information appears at distances in deep layers, while those shows in the early layers are close for speech.

% It is known that the hierarchy of receptive fields allows the CNN network to extract high-level representations such as words or faces by building from low-level features in the early layers \cite{lee2009convolutional,donahue2014decaf}.

%It also aligns with the view that the CNN networks can extract both high-level and low-level representations.
%\subsection{Localized Patterns}
% In addition, we note the offsets in \textit{Layer1} is non-trivial since the values can range in 7\%-30\% of the kernel size.

\subsection{Global Pattern}
%TODO: Mention same utterance, across utterances
%TODO: Find better color ..
%TODO: Fix caption (explain color, input, output positions)
%TODO: Is there a better plot??
%TODO: Talk more about what you observe, scores, information above diagonal, and so on ..
\vspace{-0.5em}
\begin{figure}[h!]
     \centering
     \begin{subfigure}[b]{0.23\textwidth}
         \centering
         \includegraphics[width=\textwidth]{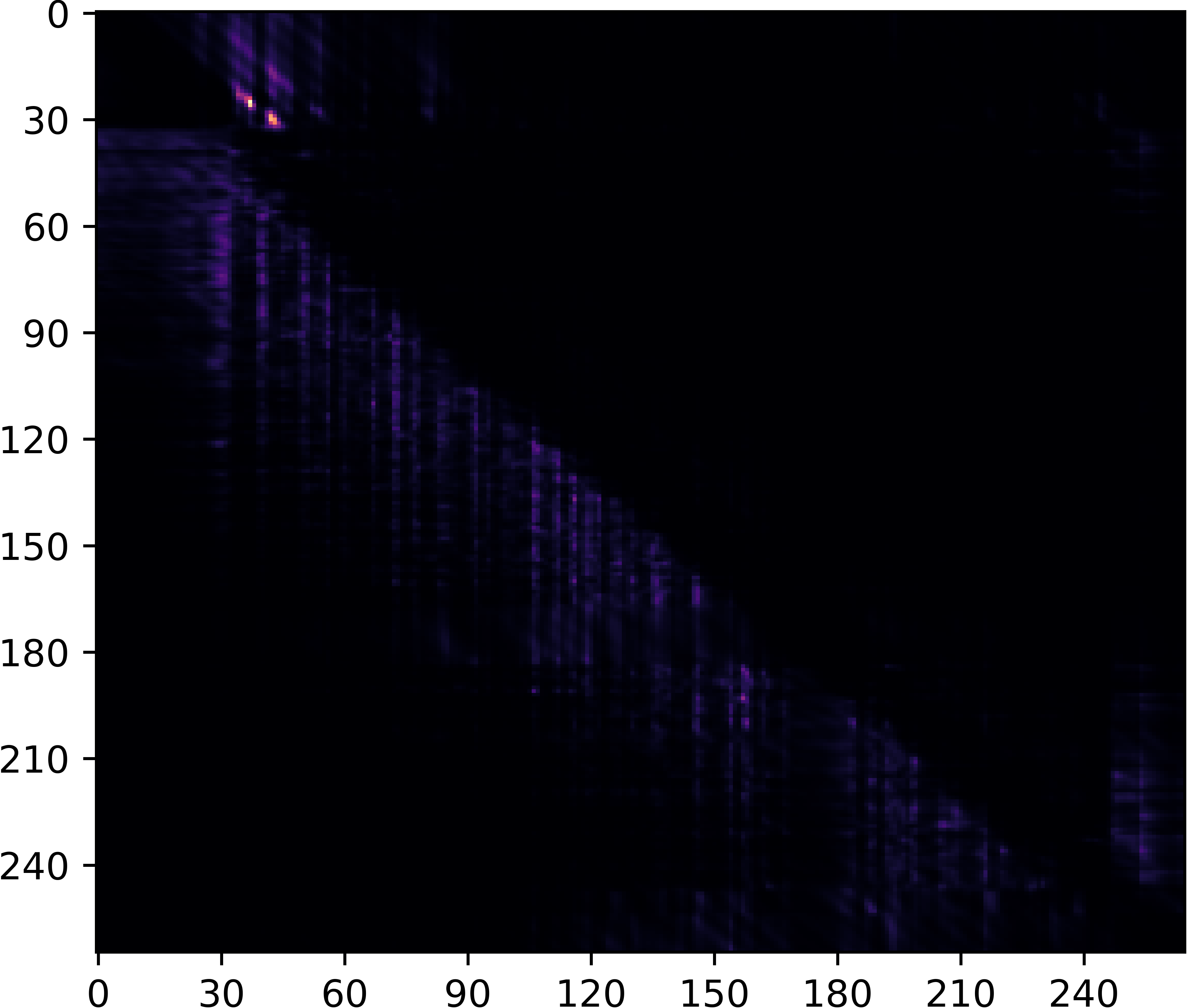}
         \caption{Conformer attention map}
         \label{ConfAtt}
     \end{subfigure}
     \begin{subfigure}[b]{0.23\textwidth}
         \centering
         \includegraphics[width=\textwidth]{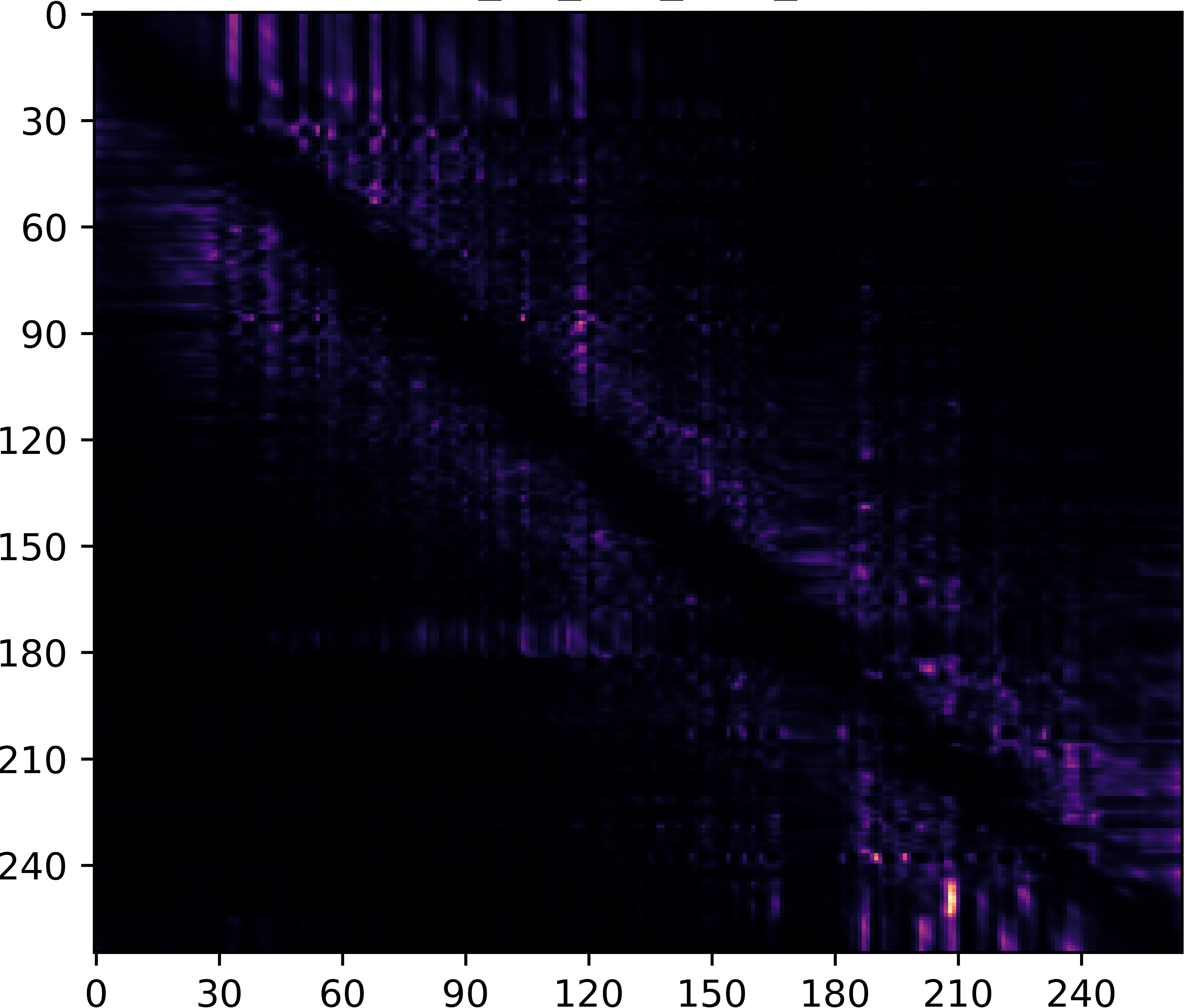}
         \caption{Deformer attention map}
         \label{DefAtt}
     \end{subfigure}
     \caption{Attention maps obtained by the last layer of the Conformer and Deformer for the 4k0c030i utterance. Output is on the vertical axis that attends to input on the horizontal axis.}
     \label{defatt}
     \vspace{-0.7em}
\end{figure}
\noindent Fig. \ref{defatt} compares attention maps obtained by the last layer of the Conformer and the Deformer encoder for a single utterance. From the graphs, we may speculate that both provide associations to the representations on the diagonal. The Conformer attends only to the past except for the first 30 locations where no history exists. But the Deformer attends additionally to the future above the diagonal. This may indicate deformed local features induce more global relevance than regular features.
% In details, the metrics evaluate an attention head for the degree of spreading along the overall, vertical, and diagonal directions.
\vspace{-0.2em}
\renewcommand{\arraystretch}{0.15}
\begin{table}[h!]
\centering
\resizebox{\columnwidth}{!}{
    \begin{tabular}{c c c c}
    \toprule
    \textbf{Model} & \makecell[c]{\textbf{Globalness}\\
    $(0, 7.48]$} & \makecell[c]{\textbf{Verticality}\\ $[ -7.48, 0)$} & \makecell[c]{\textbf{Diagonality}\\ $\left[-0.75, 0\right]$} \\
    \midrule
    Conformer & 4.57 & -6.95 & -0.18 \\
    \midrule
    Deformer & 4.82 & -6.97 & -0.19 \\
    \bottomrule
    \end{tabular}
}
\caption{Quality evaluation of the attention heads.}
\vspace{-1.5em}
\label{att_eval}
\end{table}
\renewcommand{\arraystretch}{}

To verify our observation, we evaluate the spread of attention for each head using the metrics in \cite{yang2020understanding} and report the average value across all heads on the combined set. The globalness measures the average entropy of attention distributions (an overall spread). The verticality measures the entropy of averaged attention distribution (a vertical spread). The diagonality measures the diagonal spread that helps obtain speech alignments \cite{kim2017joint} and achieves maximum at zero when the attention is on the main diagonal. For all metrics, a larger score indicates more spread. Note the bounds of all metrics depend on lengths of the encoded sequences in the corpus, which we calculate and include in Table \ref{att_eval}. It shows attention obtained by the Deformer indeed has an increase in global spread, which verifies the boost of overall feature associations led by deformation. Based on the metrics, we know that a random increase in attention globally can cause a severe decrease in both verticality and diagonality. But in the range of each respective bounds, we observe less significant decreases in both metrics compared to the increase in globalness ($-0.3\%$ or $-1.3\%$ vs. $+3.3\%$), indicating the Deformer boosts feature associations while retaining the original attention structure such as the diagonals shown in Fig. 5b.

\section{Experiments}
The experiments are conducted on the Wall Street Journal (WSJ) corpus using recipes from the Espnet toolkit \cite{watanabe2018espnet}. WSJ \cite{paul1992design} is a standard benchmark dataset to evaluate speech recognition system performance. The WSJ dataset includes read speech with transcripts drawn from the newspaper. The data is partitioned into 81 hours of training speech (\textit{si284}), 1 hour for development (\textit{dev93}), and 0.7 hour for evaluation (\textit{eval92}). The average utterance length is 7.8 seconds. Except for the SpecAug, we do not utilize any data augmentation methods. 

% The experiments are conducted on the WSJ and Switchboard (SWB) speech corpus using recipes from the Espnet toolkit \cite{watanabe2018espnet}. Both the WSJ \cite{paul1992design} and Switchboard is standard benchmark datasets to evaluate speech recognition system performance. The WSJ dataset includes read speech with transcripts drawn from the newspaper. The data is partitioned into 81 hours of training speech for (\textit{si284}), 1 hour for development (\textit{dev93}), and 0.7 hour for evaluation (\textit{eval92}). The average length of an utterance is 7.8 seconds. The SWB speech corpus contains spontaneous speech from two sides of a conversation over the telephone line. The data has 317 hours of speech in total. The average length of an utterance is 5.4 seconds. We further augment each utterance 3-fold using speed perturbation. The SWB and CHM subset of Hub5'00 are used as evaluation. 

\subsection{Baseline System}

% \section{Results}
% \subsection{Baseline}
Our baseline for WSJ follows the identical setup provided in the Espnet except for the batch size and GPUs. The model has a 12-layer Conformer encoder and a 6-layer Transformer decoder. The model dimension is 256 and has 4 attention heads. The system is trained based on English letters by the joint ctc-attention loss with a ctc weight of 0.3. The Adam \cite{kingma2014adam} optimizer is used with a warmup learning scheduler, setting $lr\_peak = 0.005$ and $warmup\_steps = 30,000$. We use two GPUs with 1.25M elements in each batch. Unless stated otherwise, these hyperparameters are kept the same in the following experiments.
\subsection{The Deformer Encoder Layer Setup}
The Deformer encoder has a default setup with deformable layers at 1, 6, 7, 10, and 11. A learning rate multiplier is commonly set for the offset CNN to learn in different rates than the rest of the model. We find both multipliers $mult=0.5$ and $1.0$ work better and report the results with these settings below.

\subsubsection{Parameter Initialization}
The parameter initialization is crucial to training the Deformer encoder. It is common to start parameters of the offset CNN from zero for the deformable CNN. But the overall parameters of the model are initialized by the Xavier \cite{glorot2010understanding} method. We test and compare these two initialization strategies under varying learning rate multipliers, denoted by 1) \textit{Xavier}: all parameters are initialized by Xavier, 2) \textit{Zero}: all parameters are initialized by Xavier except the offset CNN is initialized to zero.
\begin{table}[h]
    \vspace{-0.5em}
\centering
\resizebox{\columnwidth}{!}{
    \begin{tabular}{c c c c c c c}
    \toprule
        \multirow{2}{*}{\textbf{Model}} & \multirow{2}{*}{\textbf{\#Params}} & \multirow{2}{*}{\textbf{Initialization}} & \multicolumn{2}{c}{\textbf{WER (\%)}} & \multicolumn{2}{c}{\textbf{CER (\%)}} \\
         & & & \textbf{dev} & \textbf{eval} & \textbf{dev} & \textbf{eval} \\
        \midrule
        Conformer Base & 43.05M & Xavier & 11.2 & 8.9 & 3.9 & 3.0\\
        \midrule
        Deformer (\textit{mult=0.5}) & 43.34M & Xavier & 10.7 & \textbf{8.4} & 3.7 & \textbf{2.8} \\
        Deformer (\textit{mult=1.0}) & 43.34M & Xavier & 11.1 & 9.0 & 3.9 & 3.0 \\
        Deformer (\textit{mult=0.5}) & 43.34M & Zero & 10.8 & 8.8 & 3.7 & 3.0\\
        Deformer (\textit{mult=1.0}) & 43.34M & Zero & 10.5 & \textbf{8.4} & 3.7 & 2.9 \\

    \bottomrule
    \end{tabular}
}
\caption{WERs and CERs of using different parameter initialization schemes for the Deformer.}
\label{deformer_result}
        \vspace{-1.5em}
\end{table}

As shown in Table \ref{deformer_result}, both methods can result in the best WER of 8.4\% on the evaluation set, which improves the Conformer baseline by a +5.6\% relative. The systems use different learning rate multipliers, which indicates the architecture can be sensitive to this hyperparameter. Tying all deformable layers with a single learning rate multiplier may have magnified its impact. However, the Deformer initialized by the \textit{Zero} setup has a smaller change in WERs across different learning rate multipliers comparing the \textit{Xavier} initialization. It has also outperformed the baseline in both setups. Hence, we conclude that it is beneficial to use the \textit{Zero} setup for the Deformer.

\subsubsection{Deformable Groups}
%Run 16 or/and 4..
\vspace{-1em}
\begin{table}[h]
\centering
\resizebox{\columnwidth}{!}{
    \begin{tabular}{c c c c c c}
    \toprule
        \multirow{2}{*}{\textbf{Model}} & \multirow{2}{*}{\textbf{\begin{tabular}[c]{@{}c@{}} Deformable \\ Groups\end{tabular}}} & \multicolumn{2}{c}{\textbf{WER (\%)}} & \multicolumn{2}{c}{\textbf{CER (\%)}} \\
         & & \textbf{dev} & \textbf{eval} & \textbf{dev} & \textbf{eval} \\
        \midrule
        Deformer (\textit{mult=0.5, zero init.}) & 256 & 11.2 & 9.1 & 3.9 & 3.0\\
        Deformer (\textit{mult=1.0, zero init.}) & 256 & 11.1 & 8.9 & 3.9 & 3.0\\
        \midrule
        Deformer (\textit{mult=0.5, zero init.}) & 2 & 11.1 & 8.6 & 3.8 & 2.9\\
        Deformer (\textit{mult=1.0, zero init.}) & 2 & 10.9 & 8.6 & 3.9 & \textbf{2.8}\\
        \midrule
        Deformer (\textit{mult=0.5, zero init.}) & 1 & 10.8 & 8.8 & 3.7 & 3.0\\
        Deformer (\textit{mult=1.0, zero init.}) & 1 & 10.5 & \textbf{8.4} & 3.7 & 2.9\\
    \bottomrule
    \end{tabular}
}
\caption{WERs and CERs of using different number of offset groups}
\label{deGroups}
    \vspace{-2em}
\end{table}
\noindent As previously stated in Sec. 3.2, making the offset CNN a depthwise operation is still debatable. We experimented extensively with 1, 2, and 256 deformable groups under varying learning rate multipliers. As shown in Table \ref{deGroups}, the model performs better on both the development and evaluation set when the number of offset groups reduces. We suspect it is reasonable since the small number of offset groups should help the deformable layer discover localized patterns that are general. Moreover, the potential impact of having channel dependencies on the offset output is probably very small.

\subsubsection{Language Model}
\vspace{-1em}
\begin{table}[h!]
\centering
\resizebox{0.95\columnwidth}{!}{
    \begin{tabular}{c c c c c c}
    \toprule
     \multirow{2}{*}{\textbf{Model}} & \multicolumn{2}{c}{\textbf{WER (\%)}} & \multicolumn{2}{c}{\textbf{CER (\%)}} \\
    & \textbf{dev} & \textbf{eval} & \textbf{dev} & \textbf{eval} \\
    \midrule
    Conformer Base & 7.0 & 4.7 & 3.1 & 2.1\\
    Deformer (\textit{mult=1.0, zero init.}) & 6.7 & \textbf{4.4} & 2.9 & \textbf{2.0} \\
    \bottomrule
    \end{tabular}
}
\caption{Conformer vs. Deformer on WSJ (LM weight=1.0)}
\label{lm}
    \vspace{-1.5em}
\end{table}
\noindent Finally, we investigate whether the improvement found in the encoder can persist after adding an external language model. In detail, we use a pre-trained transformer language model and balance the decoding scores using a LM weight of 1.0. As shown in Table \ref{lm}, the Deformer outperforms the Conformer baseline across the table, with a notable +6.7\% WER improvement on the WSJ \textit{eval92} set.
\vspace{-0em}

% \subsection{Spontaneous Speech}
% We test generalization on spontaneous speech.
% % \subsubsection{D-Conv2d Subsampling Layer}
% \begin{table}[h!]
% \centering
% \resizebox{\columnwidth}{!}{
%     \begin{tabular}{c c c c c c}
%     \toprule
%         \multirow{2}{*}{\textbf{Model}} & \multirow{2}{*}{\textbf{Lr Mult.}} & \multicolumn{2}{c}{\textbf{no LM (\%)}} & \multicolumn{2}{c}{\textbf{w. LM (\%)}} \\
%          & & \textbf{clhm} & \textbf{swbd} & \textbf{clhm} & \textbf{swbd} \\
%         \midrule
%         Conformer Base & None & 14.6 & 8.0 & 13.9 & 7.5\\
%         \midrule
%         % Conformer w. D-subsample & 0.2 &  &  &  & \\
%         % Conformer w. D-subsample & 0.5 &  &  &  & \\
%         % Conformer w. D-subsample & 1.0 &  &  &  & \\
%         % Deformer w. D-subsample (\textit{mult=0.5}) & 0.2 &  &  &  & \\
%         Deformer (\textit{zero init.}) & 0.2 &  &  &  & \\
%     \bottomrule
%     \end{tabular}
% }
% \caption{Initialization methods for Deformer architecture (numbers are eval/dev)}
% \label{dsubsample}
% \end{table}
% \vspace{-1em}

% \subsection{Character/Syllable Error Analysis}

\section{Conclusions \& Future Work}
In conclusion, we present a novel encoder design for end-to-end speech recognition. It is shown that the Deformer encoder strengthens localized patterns by deforming convolutional kernels to focus on task-relevant regions, which further enhances attention mechanisms. On the WSJ evaluation dataset, the Deformer achieves a notable +6.4\% relative WER improvement over the Conformer model. We notice the system can be sensitive to the learning rate multiplier due to layer tying and verify in experiments that using zero initialization for the offset CNNs and MVN for the input alleviate such issues and increase robustness. For future work, we want to improve the model robustness more, for which the solution can lead to a larger model design and exploration for more deformable layers.
\section{Acknowledgements}
The authors would like to thank Wei Xia and Szu-Jui Chen for their meaningful discussion and suggestions on the work.

\bibliographystyle{IEEEtran}

\bibliography{mybib}

% \begin{thebibliography}{9}
% \bibitem[1]{Davis80-COP}
%   S.\ B.\ Davis and P.\ Mermelstein,
%   ``Comparison of parametric representation for monosyllabic word recognition in continuously spoken sentences,''
%   \textit{IEEE Transactions on Acoustics, Speech and Signal Processing}, vol.~28, no.~4, pp.~357--366, 1980.
% \bibitem[2]{Rabiner89-ATO}
%   L.\ R.\ Rabiner,
%   ``A tutorial on hidden Markov models and selected applications in speech recognition,''
%   \textit{Proceedings of the IEEE}, vol.~77, no.~2, pp.~257-286, 1989.
% \bibitem[3]{Hastie09-TEO}
%   T.\ Hastie, R.\ Tibshirani, and J.\ Friedman,
%   \textit{The Elements of Statistical Learning -- Data Mining, Inference, and Prediction}.
%   New York: Springer, 2009.
% \bibitem[4]{YourName17-XXX}
%   F.\ Lastname1, F.\ Lastname2, and F.\ Lastname3,
%   ``Title of your INTERSPEECH 2022 publication,''
%   in \textit{Interspeech 2022 -- 23\textsuperscript{rd} Annual Conference of the International Speech Communication Association, September 18-22, Incheon, Korea, Proceedings, Proceedings}, 2022, pp.~100--104.
%Szu-Jui Chen and Wei Xia
% \end{thebibliography}

\end{document}